\documentclass[%
prl,twocolumn,amsmath,amssymb,superscriptaddress, bibnotes]{revtex4-2}

\usepackage{graphicx}
\usepackage{bm}
\usepackage{color}

\begin{document}

\title{Enhancement of charge-neutral fermionic excitation near spin-flop transition\\ in magnetic Kondo material YbIr$_3$Si$_7$}

\author{Shunsaku~Kitagawa}

\email{kitagawa.shunsaku.8u@kyoto-u.ac.jp}
\affiliation{Department of Physics, Kyoto University, Kyoto 606-8502, Japan}

\author{Takumi~Kobayashi}
\author{Fumiya~Hori}
\author{Kenji~Ishida}
\affiliation{Department of Physics, Kyoto University, Kyoto 606-8502, Japan}

\author{Andriy~H.~Nevidomskyy}
\affiliation{Department of Physics and Astronomy, Rice University, Houston, TX 77005 USA}

\author{Long~Qian}
\affiliation{Department of Chemistry, Rice University, Houston, TX 77005 USA}

\author{Emilia~Morosan}
\affiliation{Department of Physics and Astronomy, Rice University, Houston, TX 77005 USA}
\affiliation{Department of Chemistry, Rice University, Houston, TX 77005 USA}

\date{\today}

\begin{abstract}
The new Kondo material YbIr$_3$Si$_7$, similar to other Kondo insulators, has been reported to exhibit charge-neutral fermionic excitations through measurements of specific heat and thermal conductivity at low temperatures.
We performed $^{29}$Si-NMR on YbIr$_3$Si$_7$ to investigate the magnetic response of charge-neutral fermions from a microscopic perspective.
In low magnetic fields parallel to the $c$ axis, a single NMR peak in the paramagnetic state splits into three peaks below $T_{\rm N}$.
In contrast, only a slight shift of the single NMR peak was observed in high magnetic fields.
This spectral change as a function of the $c$-axis magnetic field is interpreted as spin-flop transition, at which the magnetic moments oriented along the $c$ axis (AF-I phase) are rotated to the $ab$ plane with ferromagnetic component along the $c$-axis (AF-II phase). 
In the vicinity of the spin-flop magnetic field  $H_{\rm M}$, nuclear spin-lattice relaxation rate $1/T_1$ was found to be proportional to temperature at low temperatures, indicating the existence of charge-neutral fermions.
Furthermore, a peak of $1/T_1$ vs. the $c$-axis magnetic field suggests that the charge-neutral fermions in YbIr$_3$Si$_7$ are closely related to its magnetic properties.
Our findings shed light on the origin of charge-neutral fermions in insulators.
\end{abstract}

\maketitle
A charge-neutral fermionic excitation in strongly correlated insulators is at present among the most researched topics in condensed-matter physics.
In frustrated antiferromagnets, gapless fermionic excitations have been observed in several experiments\cite{M.Yoshida_PRL_2009,M.Yamashita_Science_2010,M.Gomilsek_PRL_2017,H.Murayama_PRR_2020,N.Li_NatCommun_2020,F.Hori_arXiv_2022}.
This gapless excitation has been discussed in terms of a spin liquid with spinon Fermi surfaces\cite{XGWen_PRB_2002,L.Balents_Nature_2010,Y.Shen_Nature_2016}.
Recently, in certain Kondo insulators, quantum oscillation, specific heat, and thermal conductivity experiments have revealed the presence of itinerant fermions with bulk nature\cite{G.Li_science_2014,T.T.Terashima_PRL_2018,M.Hartstein_NatPhys_2018,Z.Xiang_science_2018,Y.Sato_NatPhys_2019}, which contradicts the charge gap observed in transport measurements.
Although various theoretical models attempting to explain novel charge-neutral fermionic excitations have been proposed\cite{O.Erten_PRL_2017,D.Chowdhury_NatCommun_2018,P.Rao_PRB_2019,C.M.Varma_PRB_2020,Y.Tada_PRB_2020}, the origin of the charge-neutral fermions still remains unclear. 

\begin{figure*}[!tb]
\includegraphics[width=18cm,clip]{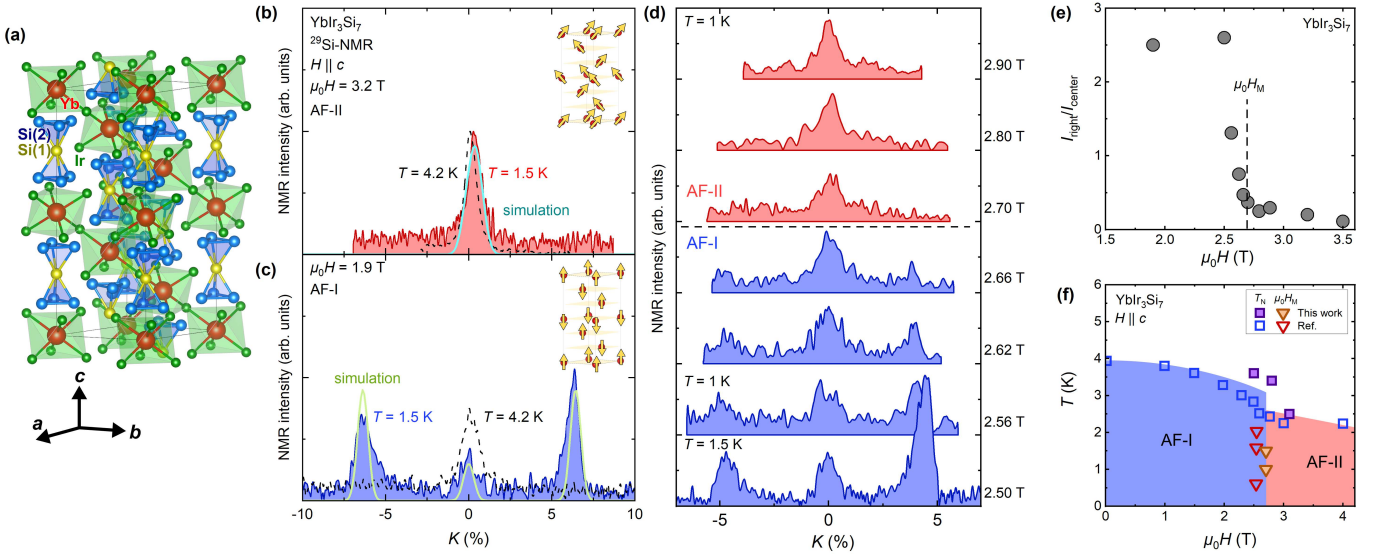}
\caption{Information about magnetic structures in YbIr$_3$Si$_7$ determined from NMR measurements.
(a) The crystal structure of YbIr$_3$Si$_7$.
$^{29}$Si-NMR spectrum at 1.5 and 4.2~K for (b) 3.2~T (AF-II phase) and (c) 1.9~T (AF-I phase).
The simulations of the NMR spectrum considering the classical dipole interaction are also plotted. 
The magnetic structures used in the simulation are shown in the inset.
(d) Magnetic field variation of NMR spectrum in the AF state.
(e) Magnetic field dependence of the intensity ratio of center and left peaks.
The dashed line indicate the spin-flop transition field $H_{\rm M}$.
(f) $H$--$T$ phase diagram for $H \parallel c$.
Squares and inverted triangles denote $T_{\rm N}$ and $H_{\rm M}$, respectively. 
}
\label{Fig.1}
\end{figure*}

Quite recently, a new class of Kondo material YbIr$_3$Si$_7$, showing insulating resistivity, has been discovered\cite{M.Stavinoha_arXiv_2021}.
YbIr$_3$Si$_7$ has a trigonal ScRh$_3$Si$_7$-type crystal structure with space group $R$\={3}$c$ (\#167, $D_{3d}^{6}$) [Fig.~\ref{Fig.1}(a)].
Magnetization and X-ray photoelectron spectroscopy (XPS) measurements indicated that the Yb-ions were very close to the trivalent state in the bulk.
At zero magnetic field, antiferromagnetic (AF) order occurs below the N\'{e}el temperature $T_{\rm N}$ = 4.1~K.
Further, neutron diffraction measurements have reported that $\Gamma_1$ AF state is realized below $T_{\rm N}$\cite{M.Stavinoha_arXiv_2021}.
In the $\Gamma_1$ AF state, all the Yb$^{3+}$ moments are oriented along the $c$ axis, with each Yb$^{3+}$ moment aligned anti-parallel to its six nearest neighbors in the nearly cubic Yb sublattice, and parallel to its co-planar next nearest neighbors, as shown in the inset of Fig.~\ref{Fig.1}(c).
The size of the ordered moment is $\sim$1.5 $\mu_{\rm B}$/Yb$^{3+}$. 
Some of the authors previously reported that this AF order can be tuned by the magnetic field parallel to the $c$ axis and that a spin-flop transition, at which the magnetic moments oriented along the $c$ axis (AF-I phase) are rotated to the $ab$ plane (AF-II phase), occurs at $\mu_0H \sim 2.5$~T [Fig.~\ref{Fig.1}(f)]\cite{Y.Sato_arXiv_2021}.
Moreover, the low-temperature specific heat and thermal conductivity measurements in YbIr$_3$Si$_7$ suggested the presence of charge-neutral fermionic excitations\cite{Y.Sato_arXiv_2021}.

In this study, we performed $^{29}$Si-NMR measurements on YbIr$_3$Si$_7$ to investigate the magnetic properties at low temperatures as well as the relation with the charge-neutral fermions from a microscopic point of view.
The obtained NMR results support the existence of charge-neutral fermionic excitations and indicate that the charge-neutral fermions are closely related to the spin-flop transition in YbIr$_3$Si$_7$.

High-quality single crystals of YbIr$_3$Si$_7$ were grown as described in ref.~\cite{M.Stavinoha_arXiv_2021}.
There are two crystallographically inequivalent Si sites, 6a and 36f, with point symmetries 32 and 1, which are denoted as Si(1) and Si(2), respectively, as shown in Fig.~\ref{Fig.1}(a).
A conventional spin-echo technique was used for NMR measurements.
$^{29}$Si ($I = 1/2$, nuclear gyromagnetic ratio $^{29}\gamma_n/2\pi = 8.4587$~MHz/T, and natural abundance 4.7\%) NMR spectra were obtained as a function of frequency at fixed magnetic fields. 
The NMR spectra are plotted against $K \equiv (f-f_0)/f_0$, where $f$ is the NMR frequency, and $f_0$ is the reference frequency determined as $f_0 = (\gamma_n/2\pi)\mu_0H$. 
The magnetic field was calibrated using $^{63}$Cu ($^{63}\gamma_n/2\pi = 11.285$~MHz/T) and $^{65}$Cu ($^{65}\gamma_n/2\pi = 12.089$~MHz/T) NMR signals from the NMR coil. 
The nuclear spin-lattice relaxation rate $1/T_1$ was determined by fitting the time variation of the nuclear magnetization after saturation probed with the spin-echo intensity to a theoretical function for $I$ = 1/2.
Low-temperature NMR measurements down to 110~mK were performed using an $^{3}$He-$^{4}$He dilution refrigerator, and the single-crystalline sample was immersed into the mixture.

First, we investigated the $c$-axis magnetic field variation of the magnetic structure.
Figures~\ref{Fig.1}(b) and \ref{Fig.1}(c) show $^{29}$Si-NMR spectrum for 3.2~T (AF-II phase) and 1.9~T(AF-I phase), respectively.
At 4.2~K ($> T_{\rm N}$), a single NMR peak was observed both at 1.9 and 3.2~T, reflecting the overlapping of two Si sites.
As reported previously\cite{Y.Sato_arXiv_2021}, at 1.9~T (AF-I phase), the NMR spectrum splits into three peaks below $T_{\rm N}$, with the peaks in the higher and lower $K$, indicating the appearance of an internal magnetic field parallel to the external magnetic field at the Si(2) site.
The remaining peak arises from the Si(1) site at which an internal magnetic field from the Yb magnetic moment is canceled. 
Considering the classical dipole interaction, these three peaks can be reproduced by the $\Gamma_1$ AF state with 1.5 $\mu_{\rm B}$/Yb$^{3+}$, as shown in the inset of Fig.~\ref{Fig.1}(c).
Here, the dipolar magnetic field at each Si site $\bm{H}_{\rm int}^{\rm Si}$ from the Yb 4$f$ moments can be expressed as,
\begin{align}
\bm{H}_{\rm int}^{\rm Si} &= \sum_i \bm{H}^{\rm dip}_i (\bm{r}_i),\\
\bm{H}^{\rm dip}_i(\bm{r}_i) &= -\nabla \frac{\bm{\mu}_i \cdot \bm{r_i}}{4\pi \bm{r}_i^{3}},
\end{align}
where $\bm{\mu}_i$ is a magnetic moment at the $i$th Yb site, and $\bm{r}_i$ is the relative position vector of $i$th Yb site from the Si site.
The magnetic structure and size of ordered moment are consistent with neutron diffraction measurements\cite{M.Stavinoha_arXiv_2021}.
On the other hand, only a slight shift to higher $K$ direction was observed at 3.2~T (AF-II phase).
This small shift can be reproduced via the magnetic structure as shown in the inset of Fig.~\ref{Fig.1}(b) with 1.5 $\mu_{\rm B}$/Yb$^{3+}$, wherein, the AFM moments are oriented perpendicular to the external magnetic field with ferromagnetic component along the field.
The angle from the $ab$ plane was estimated to be approximately 75$^{\rm o}$ (almost parallel to the $c$ axis).
In this magnetic structure, the internal fields at the Si(1) and Si(2) sites are small and almost identical, thereby resulting in the single $^{29}$Si-NMR peak even below $T_{\rm N}$.
Consequently, these results indicate the occurrence of a spin-flop transition with applying magnetic field along the $c$ axis. 

To determine the spin-flop transition field $H_{\rm M}$ from NMR measurements, we measured the $c$-axis magnetic field variations of NMR spectrum below $T_{\rm N}$ as shown in Fig.~\ref{Fig.1}(d).
Up to $\sim$ 2.66~T, three distinct peaks were observed, indicating the AF-I phase.
With the further application of a magnetic field, the NMR spectrum reduced to a single peak, reflecting the spin-flop transition.
The change in the NMR spectrum can also be recognized in the magnetic field dependence of the intensity ratio of center and right peaks as shown in Fig.~\ref{Fig.1}(e).
The intensity of the right peak begins to decrease from 2.5~T and attained a background level at 2.7~T with increasing magnetic field.
Thus, $H_{\rm M}$ was determined to be $\sim$2.7~T, which is in good agreement with the magnetization results\cite{Y.Sato_arXiv_2021}, as shown in Fig.~\ref{Fig.1}(f).

\begin{figure}[!tb]
\includegraphics[width=8.5cm,clip]{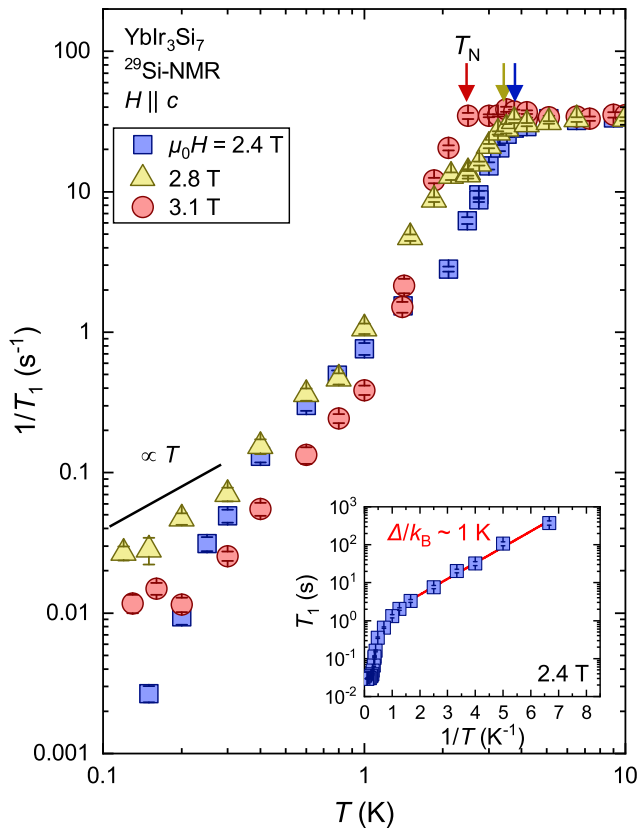}
\caption{
Temperature dependence of nuclear spin-lattice relaxation rate $1/T_1$ for $\mu_0H$ = 2.4, 2.8, and 3.1~T for $H \parallel c$.
The solid line is a guide for the eye of temperature-linear behavior.
Inset: Inverted temperature dependence of $T_1$ at 2.4~T.
Considering the low-temperature behavior, the magnetic excitation gap was estimated to be $\sim$ 1~K.
The solid line is the fitting result.
}
\label{Fig.2}
\end{figure}

\begin{figure}[!tb]
\includegraphics[width=8.5cm,clip]{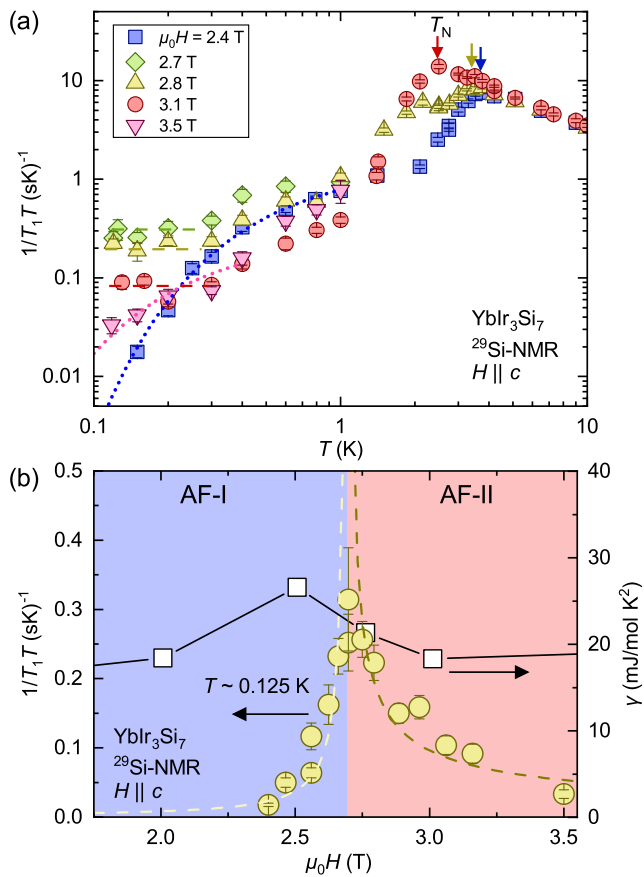}
\caption{
(a) Temperature dependence of $1/T_1T$ for various $c$-axis magnetic fields.
The dotted curves are guides for the eye of gaped behavior.
(b) Magnetic field dependence of $1/T_1T$ at $\sim$ 0.125~K.
For comparison, we also plot the magnetic field dependence of the electronic specific heat coefficient $\gamma$\cite{Y.Sato_arXiv_2021}.
}
\label{Fig.3}
\end{figure}

Next, we investigate the low-energy magnetic excitations with respect to the $c$-axis magnetic field.
Figure~\ref{Fig.2} shows the temperature dependence of the nuclear spin-lattice relaxation rate $1/T_1$ for $\mu_0H$ = 2.4, 2.8, and 3.1~T ($H \parallel c$).
In the measured magnetic field range, $1/T_1$ was approximately constant above $T_{\rm N}$, indicating that AF fluctuations are governed by localized moments.
$T_{\rm N}$ was determined by the temperature below which $1/T_1$ suddenly decreased.
$T_{\rm N}$ determined via the NMR measurements in this study was slightly higher than that determined from bulk measurements [Fig.~\ref{Fig.1}(f)].
In contrast to the behavior in high temperatures, the low-temperature behavior was largely affected by magnetic fields.
Below $T_{\rm N}$, $1/T_1T$ decreased on cooling owing to the suppression of magnetic fluctuations.
At 2.4~T (AF-I phase), $1/T_1$ decreased exponentially on cooling, indicating the existence of a spin gap.
Such behavior is common in insulating magnets.
The size of the gap was estimated to be $\sim$ 1~K from the Arrhenius plot as shown in the inset of Fig.~\ref{Fig.2}.
On the other hand, at 2.8 and 3.1~T, $1/T_1$ is proportional to temperature at low temperatures, indicating the non-zero density of states at Fermi energy $E_{\rm F}$.
In metallic systems, nonzero $1/T_1T$ is observed because of the contribution from conduction electrons; however, its origin in case of insulators is not determined so far.
Such nonzero values of $1/T_1T$ in the AF-II phase were clearly observed from the residual $1/T_1T$ via extrapolation to $T \rightarrow 0$, as in Fig.~\ref{Fig.3}(a).
The constant value at low temperatures decreased with increasing magnetic field, and, at 3.5~T, it becomes approximately one order of magnitude smaller than the maximum value.
Figure~\ref{Fig.3} (b) depicts the magnetic field dependence of $1/T_1T$ at $\sim$ 0.125~K.
$1/T_1T$ shows a peak at 2.7~T, which coincidences with $H_{\rm M}$.
The significant enhancement of 1/$T_1T$ near $H_{\rm M}$ has also been observed in conventional antiferromagnets due to higher-order magnon processes\cite{D.Paquette_PRB_1975,A.J.Wal_PRB_1981}.
However, for higher-order magnon processes, the temperature dependence of 1/$T_1$ decreases exponentially\cite{T.Moriya_PTEP_1956,N.Kaplan_PRL_1966}.
In YbIr$_3$Si$_7$, the temperature dependence of 1/$T_1$ at low temperatures is proportional to $T$.
Thus, the observed behavior cannot be explained by conventional magnon processes.
We fit the magnetic field dependence of $1/T_1T$ to $|H_{\rm M} - H_0|^{- \beta}$, as shown in Fig.~\ref{Fig.3}~(b).
The $\beta$ is 1.2 below $H_{\rm M}$ and 0.6 above $H_{\rm M}$, which is smaller than conventional antiferromagnets (1.5)\cite{D.Paquette_PRB_1975,A.J.Wal_PRB_1981}.

The nonzero $1/T_1T$ at low temperatures in the insulating state has been observed in frustrated Kagom\'{e} compounds, such as Cu$_3$V$_2$O$_7$(OH)$_2 \cdot$2H$_2$O\cite{M.Yoshida_PRL_2009} and ZnCu$_3$(OH)$_6$SO$_4$\cite{M.Gomilsek_PRL_2017}.
Furthermore, recently, some of the authors reported that constant $1/T_1T$ behavior in the AF state of the Yb-based semiconductor YbCuS$_2$ with 4$f$ zigzag structure.
The value of $1/T_1T$ in YbCuS$_2$ is one order of magnitude larger than that of Cu metal\cite{F.Hori_arXiv_2022}.
In these systems, spinon excitation is one of the leading candidates for the origin, however, it has not yet been settled.
In contrast, a broad peak with respect to temperature has been observed in $1/T_1$ of $^{11}$B-NMR on nonmagnetic Kondo insulators SmB$_6$\cite{M.Takigawa_JPSJ_1981,P.Schlottmann_PRB_2014} and YbB$_{12}$\cite{K.Ikushima_PhysicaB_2000}, which is different from the results of YbIr$_3$Si$_7$.
Thus, further research is necessary for a unified understanding of charge-neutral fermions. 

Finally, the NMR results obtained in this study were compared with the recent results of specific heat and thermal conductivity\cite{Y.Sato_arXiv_2021}.
As shown in Fig.~\ref{Fig.3}(b), similar to $1/T_1T$, the magnetic field dependence of the electronic specific heat coefficient $\gamma$ also yielded a broad maximum at approximately $H_{\rm M}$.
However, the height and width are different between two measurements.
The peak of 1/$T_1T$ is sharper and larger than that from specific heat, which can be attributed to the sensitivity of the detection of the charge-neutral fermions at low temperatures.
Because the nuclear Schottky contribution becomes dominant at low temperatures in the specific heat measurements, the contribution of charge-neutral fermions can only be estimated by the extrapolation of $C(T)/T$ above 1~K to $T = 0$\cite{Y.Sato_arXiv_2021}.
However, there is no such background contribution in $1/T_1$; it can directly detect the contribution of charge-neutral fermions down to the lowest temperatures.
Therefore, it is considered that the charge-neutral fermionic excitations are strongly enhanced only in the narrow region around $H_{\rm M}$.

In contrast, the magnetic field dependence of thermal conductivity does not show any peak at $H_{\rm M}$\cite{Y.Sato_arXiv_2021}.
This difference originates from the characteristics of the measurement method.
In metallic systems, the specific heat and NMR-$1/T_1$ are dominated by the heavy-electron bands, whereas thermal conductivity is governed by light-electron bands.
Thus, the enhancement of the excitations around $H_{\rm M}$ is attributed to heavier (conductive) fermions.
In addition, the different behavior between NMR-$1/T_1$ and thermal conductivity suggests the existence of at least two types of charge-neutral fermions in YbIr$_3$Si$_7$.
There are many theories on the origin of charge-neutral fermions in YbIr$_3$Si$_7$, such as 3D Majorana fermions, composite magnetoexcitons, and spinons in fractionalized Fermi liquids\cite{O.Erten_PRL_2017,D.Chowdhury_NatCommun_2018,P.Rao_PRB_2019,C.M.Varma_PRB_2020,Y.Tada_PRB_2020}, but few experiments on their properties.
Our experimental results would largely contribute to the understanding of the origin.

In conclusion, we have performed $^{29}$Si-NMR on the AF Kondo material YbIr$_3$Si$_7$ to investigate the magnetic response of charge-neutral fermions from a microscopic point of view.
Based on the $c$-axis magnetic field variation of the NMR spectrum, we confirmed the spin-flop transition at $H_{\rm M} = 2.7$~T.
While $1/T_1$ decreases exponentially at low temperatures at 2.4~T, $1/T_1$ is proportional to temperature at low temperatures at 2.8, and 3.1~T, indicating the existence of charge-neutral fermions.
The peak of $1/T_1$ with respect to the $c$-axis magnetic field indicates that these charge-neutral fermions are closely related to the magnetic properties of the metamagnetic transition in YbIr$_3$Si$_7$.
Therefore, the present findings provide significant information regarding the origin of charge-neutral fermions recently discovered in Kondo materials.

\section*{Acknowledgments}
The authors thank S. Suetsugu, Y. Kasahara, Y. Matsuda, S. Yonezawa, and Y. Maeno for valuable discussions.
This work was supported by Kyoto Univ. LTM center. 
We are also supported by Grants-in-Aid for Scientific Research (KAKENHI) (Grants No. JP20H00130, No. JP20KK0061, No. JP19H04696, No. JP19K14657, and No. 21K18600).
E. M. and L. Q. acknowledge support from U.S. DOE BES DE- SC0019503.

\end{document}